# Thermodynamics of a Modified Fermi-Hubbard Model

Moorad Alexanian

*Department of Physics and Physical Oceanography*
*University of North Carolina Wilmington, Wilmington, NC 28403-5606*

E-mail: alexanian@uncw.edu

**Abstract.** A recently introduced recurrence-relation ansatz applied to the Fermi-Hubbard model gives rise to a soluble model and here is used to calculate several thermodynamic observables. The constraint of unit density per site, $\rho = 1$, is applied and some of the results are compared to cases where the constraint is not imposed. The modified model exhibits a continuous phase transition (second order) reminiscent of the integer quantum Hall resistance and a ground state, first-order phase transition.

## 1. Introduction

The Fermi-Hubbard model has become the basis for much of what we know about superfluidity in periodic systems, quantum magnetism, and strongly correlated fermion physics on lattices in general [1]. We consider a recently introduced recurrence-relation ansatz for the hopping part of the Fermi-Hubbard model giving rise to an exactly soluble [2]. We use the latter model in this work to calculate several thermodynamic observables and consider possible universality viz-à-viz the number $N$ of distinguishable microstates [3].

This paper is structured as follows. In Sec. 2, we present the Fermi-Hubbard model on an infinite, one-dimensional lattice. In Sec. 3, we introduce a recurrence-relation ansatz for the external degree of freedom associated with the nearest neighbor of the $j$-th lattice site. In Sec. 4, the grand canonical partition function is determined. In Sec. 5, the density, number of on-site pairs, compressibility, etc. are calculated. In Sec. 6, the total energy, specific heat, and entropy are determined. Finally, Sec. 8 summarizes our results.

## 2. Fermi-Hubbard model

The SU(N) Fermi-Hubbard model is given by [4, 5]

$$\hat{H} = -t \sum_{\langle i,j\rangle,\sigma} (\hat{c}^\dagger_{i\sigma}\hat{c}_{j\sigma} + \hat{c}^\dagger_{j\sigma}\hat{c}_{i\sigma}) + \frac{U}{2} \sum_{i,\sigma\neq\tau} \hat{n}_{i\sigma}\hat{n}_{i\tau} - \mu \sum_{i,\sigma} \hat{n}_{i\sigma}, \qquad (1)$$

where $\hat{c}^\dagger_{i\sigma}$, $\hat{c}_{i\sigma}$, $\sigma \in \{1...N\}$, $\hat{n}_{i\sigma} = \hat{c}^\dagger_{i\sigma}\hat{c}_{i\sigma}$ and represent the fermionic creation and annihilation operators at site $i$ with spin is the number operator, $\langle i, j\rangle$ denotes adjacent sites on a rectangular lattice, $t$ is the hopping amplitude, $U$ is the on-site interaction strength, and $\mu$ denotes the chemical potential. Representing the Fermi-Hubbard Hamiltonian on a quantum computer requires a fermionic encoding. The well-known Jordan-Wigner transform is used, under which each fermionic mode maps to one qubit, interpreted as lying on a 1D line [4]. Since we are interested in comparing our numerical results to those where universality in the number of distinguishable microstates N is considered, we are assuming that all sites have the same chemical potential $\mu$ [3].

## 3. Ansatz

Consider the following recurrence-relation ansatz [2] for the term associated with the hopping term of the *i*-th lattice in (1)

$$\hat{c}_{i+1\sigma} = \hat{c}_{i\sigma} - \hat{c}_{i-1\sigma}. \tag{2}$$

and so

$$\sum_{i,\sigma} \left[ \hat{c}^\dagger_{i\sigma}\hat{c}_{i+1\sigma} + \hat{c}^\dagger_{i+1\sigma}\hat{c}_{i\sigma} + \hat{c}^\dagger_{i\sigma}\hat{c}_{i-1\sigma} + \hat{c}^\dagger_{i-1\sigma}\hat{c}_{i\sigma} \right] = 2\sum_{i,\sigma} \hat{c}^\dagger_{i\sigma}\hat{c}_{i\sigma} = 2\sum_i \hat{n}_i, \tag{3}$$

$\hat{n}_i = \sum_\sigma \hat{n}_{i\sigma}.$

where It is clear from (3) that we are considering a truly infinite one-dimensional lattice rather than an open-ended infinite chain.

More generally on may consider the ansatz that applies to both the infinite lattice, as well as, the open-ended infinite chain, viz.,

$$\hat{c}^\dagger_{i+1\sigma} = (\alpha + i\beta)\hat{c}^\dagger_{i\sigma}, \tag{4}$$

with $\alpha$ and $\beta$ real. Therefore,

$$\sum_{i=-\infty,\sigma}^{\infty} \left[ \hat{c}^\dagger_{i\sigma}\hat{c}_{i+1\sigma} + \hat{c}^\dagger_{i+1\sigma}\hat{c}_{i\sigma} + \hat{c}^\dagger_{i\sigma}\hat{c}_{i-1\sigma} + \hat{c}^\dagger_{i-1\sigma}\hat{c}_{i\sigma} \right] = 4\alpha \sum_{i=-\infty,\sigma}^{\infty} \hat{c}^\dagger_{i\sigma}\hat{c}_{i\sigma} = 4\alpha \sum_{i=-\infty}^{\infty} \hat{n}_i \tag{5}$$

and

$$\sum_{i=1,\sigma}^{\infty} \left[ \hat{c}^\dagger_{i\sigma}\hat{c}_{i+1\sigma} + \hat{c}^\dagger_{i+1\sigma}\hat{c}_{i\sigma} \right] = 2\alpha \sum_{i=1,\sigma}^{\infty} \hat{c}^\dagger_{i\sigma}\hat{c}_{i\sigma} = 2\alpha \sum_{i=1}^{\infty} \hat{n}_i. \tag{6}$$

The left-hand-side of (4) obeys the same anticommutation relations as the right-hand-side of (4) provided $\alpha^2 + \beta^2 = 1$. In what follows, we will consider the infinite lattice with ansatz (2), that is, $\alpha = 1/2$ in (5), which is the same as for the open-ended infinite chain but with $\alpha = 1$ in (6).

The Fermi-Hubbard model (1) is reduced to

$$\begin{aligned}\hat{H} &= -2t\sum_i \hat{n}_i + \frac{U}{2}\sum_{i,\sigma\neq\tau} \hat{n}_{i\sigma}\hat{n}_{i\tau} - \mu\sum_{i,\sigma} \hat{n}_{i\sigma}, \\ &= -2t\sum_i \hat{n}_i + \frac{U}{2}\sum_i \hat{n}_i^2 - \frac{U}{2}\sum_i \hat{n}_i - \mu\sum_i \hat{n}_i,\end{aligned} \tag{7}$$

$\hat{n}_{i\sigma}^2 = \hat{n}_{i\sigma}$

since with energy eigenvalues

$$E_i = -2tn_i + \frac{U}{2}n_i^2 - \frac{U}{2}n_i - \mu n_i. \tag{8}$$

## 4. Grand canonical partition function

The grand canonical partition function is the following product over the differing lattices sites

$$\mathscr{Z} = \prod_{i,\sigma} e^{-\beta E_i}, \qquad (9)$$

where $\beta = 1/k_B T$ and we consider $N_s$ lattice sites with each containing up to $N$ distinguishable microstates $\sigma$, viz., and $\sum_i n_i = N$ so $\sum_i \langle \hat{n}_i \rangle = N$. One obtains, with the aid of (8),

$$\mathscr{Z} = \prod_i \sum_{n=0}^{N} \frac{N!}{n!(N-n)!} e^{\left[(2t/U+1/2+\mu/U)n - n^2/2\right]U/(k_B T)} = \prod_i \sum_{n=0}^{N} \frac{N!}{n!(N-n)!} e^{(\tilde{\mu} n - n^2/2)/\tilde{T}}, \qquad (10)$$

where the renormalized chemical potential $\tilde{\mu}_i$ and the scaled temperature $\tilde{T}$ are given by [2]

$$\tilde{\mu} = 2t/U + 1/2 + \mu/U \quad \text{and} \quad \tilde{T} = k_B T/U \qquad (11)$$

and the binomial coefficient $\frac{N!}{n!(N-n)!}$ represents the distribution of $n$, $n = 0 \cdots N$, distinguishable microstates in the $i$-th, $i = 1 \ldots N$, distinguishable lattice sites with all parameters expressed in units of $U$. Accordingly, there are only the variables $N$, $\tilde{\mu}$ for each lattice site $i$, and $\tilde{T}$.

It is important to remark that a given value of the renormalized chemical potential $\tilde{\mu}$ does not determine the individual values of either $t/U$ or $\mu/U$. It is clear that our modified Fermi-Hubbard model reduces to the original Fermi-Hubbard model for $t = 0$. It may be that the results from the modified Fermi-Hubbard model for $t > 0$ and $\mu \gg t$ reproduce those of the original Fermi-Hubbard model. In what follows, we calculate various thermodynamic observables for the modified Fermi-Hubbard model given by the grand partition function (10).

## 5. Density $\rho$, number of on-site pairs $D$, compressibility $\kappa$, dependence on chemical potential $\mu/t$, and $dD/d\rho$ on $\rho$

The average number $\langle \hat{n} \rangle$ of microstates in a lattice site with chemical potential $\tilde{\mu}$ and temperature $\tilde{T}$ is given by

$$\rho \equiv \langle \hat{n} \rangle = k_B T \frac{\ln \mathscr{Z}}{\partial \mu} = \frac{\sum_{n=0}^{N} \frac{N!}{n!(N-n)!} n e^{(\tilde{\mu} n - n^2/2)/\tilde{T}}}{\sum_{n=0}^{N} \frac{N!}{n!(N-n)!} e^{(\tilde{\mu} n - n^2/2)/\tilde{T}}}. \qquad (12)$$

Fig.1(a) shows the density dependence on chemical potential $(\mu - \mu_0)/t$ for $N = 2, 3, 4$, $U/t = 8$ at $T/t = 0.5$ and where $\mu_0$ is defined by $\rho(\mu_0) = 1$. Note that the universality in $N$ holds for $-5 < (\mu - \mu_0)/t < 10$, which is a larger region than that considered in [3], viz., $-5 < (\mu - \mu_0)/t < 5$. The existence of the phase transition is evident for large values of $(\mu - \mu_0)/t$ via the corresponding steps as indicated previously [2].

The number of on-site pairs per site is

$$D \equiv \frac{1}{N_s} \sum_i \left[ \frac{1}{2} \sum_{\alpha \neq \tau} \langle \hat{n}_{i\sigma} \hat{n}_{i\tau} \rangle \right] = \frac{1}{2 N_s} \sum_i \left[ \langle \hat{n}_i^2 \rangle - \langle \hat{n}_i \rangle \right] = \frac{1}{2}(\langle \hat{n}^2 \rangle - \langle \hat{n} \rangle), \qquad (13)$$

where the last equality follows since all the lattice sites are identical, that is, have the same chemical potential $\mu$. The isothermal compressibility is defied by

$$\kappa_T \equiv \left( \frac{\partial \langle \hat{n} \rangle}{\partial \mu} \right)_T = T^{-1}(\langle \hat{n}^2 \rangle - \langle \hat{n} \rangle^2), \qquad (14)$$

where $N_s$ is the number of sites and $\hat{n}_i^2 = \hat{n}_i$ for fermions. Fig.1(b) is a plot of the number of on-site pairs $D$ versus $(\mu - \mu_0)/t$ for $N = 2, 3, 4$, $U/t = 8$ at $T/t = 0.5$ and where $\mu_0$ is defined by $\rho(\mu_0) = 1$. Note again where universality, that is, independence of $N$, occurs in a larger region as in Fig.1(a) than that obtained in [3]. The asymptotic value is given by $N(N - 1)/2$. Figs.1(c, d) are plots of the isothermal compressibility $\kappa$ and $dD/d\rho$, for the same values as those in Figs.1(a, b). The universality in the isothermal compressibility $\kappa$ is the same as that in Figs.1(a, b) and the range where $\kappa$ is periodic increasing with increasing values of $N$. However, universality in $dD/d\rho$ as shown in Fig.1(d) is limited to $\rho < 1$ with step function behavior.

The asymptotes for $\rho$ and $D$ are reached for $(\mu - \mu_0)/t \approx 5(N = 2)$, $15(N = 3)$, $25(N = 4)$ as seen in Figs.1(a, b). Whereas the stepwise asymptotic behavior in Fig.1(d) represents actually a continuous phase transition (second order), which disappears at higher temperatures [2]. The stepwise behavior is reminiscent of the integer quantum Hall resistance [6, 7].

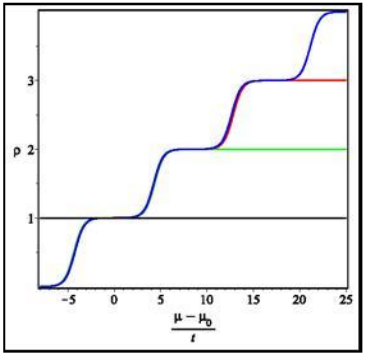

**Fig. 1a.** Density $\rho$ dependence on chemical potential, where $\rho(\mu_0) = 1$, N=2 (green), 3 (red), 4 (blue). $U/t = 8$ and $T/t = 0.5$.

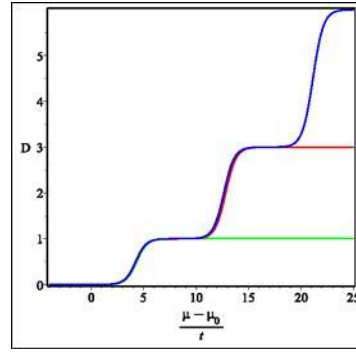

**Fig. 1b.** Number of on-site pairs dependence $D$ on chemical potential, where $\rho(\mu_0) = 1$, N=2 (green), 3 (red), 4 (blue). $U/t = 8$ and $T/t = 0.5$.

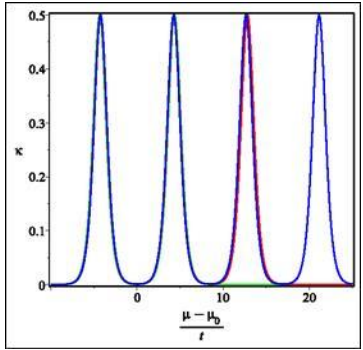

**Fig. 1c.** Compressibility $\kappa$ dependence on chemical potential, where $\rho(\mu_0) = 1$, N=2 (green), 3 (red), 4 (blue). $U/t = 8$ and $T/t = 0.5$.

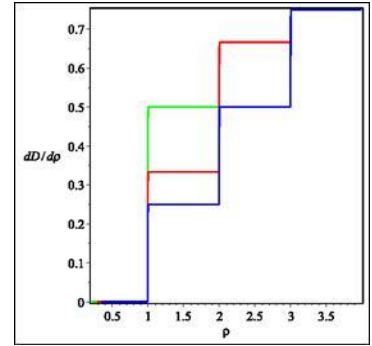

**Fig. 1d.** Derivative of the number of on-site pairs $D$ with respect to the density $dD/d\rho$ as a function of density, where $\rho(\mu_0) = 1$, N=2 (green), 3 (red), 4 (blue). $U/t = 8$ and $T/t = 0.5$.

Fig.2(a-d) show the behavior of the number of on-site pairs $D$ as a function of temperature $T/t$. Each panel compares $D$ for $N = 2, 3, 4$ for fixed values of $U/t$ and $\rho = 1$. Universality occurs for large values of $U/t$ when our modified model approaches the results of the Fermi-Hubbard model. We were not able to include the case for $N = 6$, which gives rise to a fifth order polynomial and no closed-form solutions exist for general fifth or higher order polynomial equations. In Fig.3(a-d), each panel compares $D$ for $U/t = 4, 8, 12, 40$ for fixed $N = 2, 3, 4$. Universality seems to occur for small values of $T/t$ since one has that as $T \to \infty$, $D \to 1/4, 1/3, 3/8$ for $N = 2, 3, 4$, respectively.

## 6. Total energy, specific heat, entropy

The energy *E*, the entropy *S*, and the specific heart *C* follow from the grand canonical partition function $\mathcal{Z}$

$$E = T^2\frac{\partial \ln \mathcal{Z}}{\partial T} + \mu T\frac{\partial \ln \mathcal{Z}}{\partial \mu}, \quad C = \frac{\partial E}{\partial T}, \quad \text{and} \quad S = \ln \mathcal{Z} + T\frac{\partial \ln \mathcal{Z}}{\partial T}, \tag{15}$$

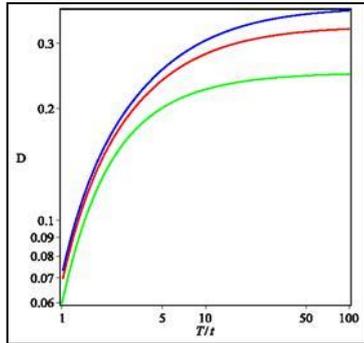

**Fig. 2a.** Number of on-site pairs *D* versus temperature for *N* = 2 (green), 3 (red), 4 (blue) for *U/t* = 4 and *ρ* = 1.

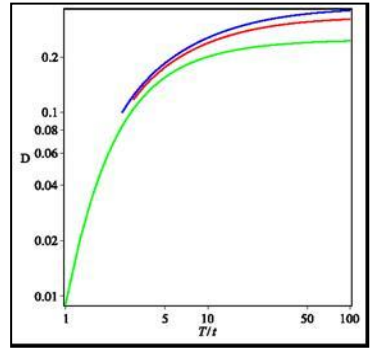

**Fig. 2b.** Same as Fig.2(a) but with *U/t* = 8.

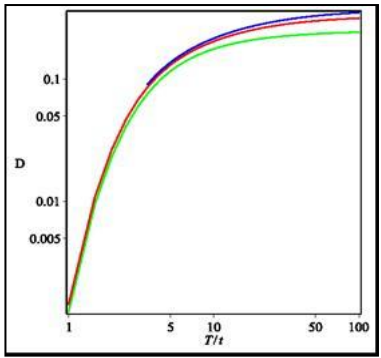

**Fig. 2c.** Same as Fig.2(a) but with *U/t* = 12.

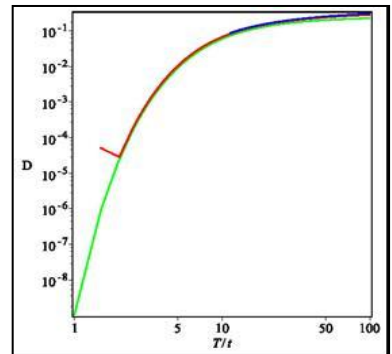

**Fig. 2d.** Same as Fig.2(a) but with *U/t* = 40.

where we have set $k_B = 1$. We shall consider, in evaluating these thermodynamic observables, the cases of lattices with *N* = 2, 3, 4 and equal chemical potentials *μ*.

Fig.4(a-d) shows the energy per site for different values of *U/t*. One has universality in a larger region than previously 0 ≤ *T/t* < 5. Actually, this behavior follows from the requirement that *ρ* = 1 since as *T* → 0, *E* → *K* and the kinetic energy per site *K* is

$$K/t = -\frac{1}{N_s}\sum_{\langle i,j\rangle,\sigma}\langle \hat{c}^\dagger_{i\sigma}\hat{c}_{j\sigma} + \hat{c}^\dagger_{j\sigma}\hat{c}_{i\sigma}\rangle = -2\rho, \tag{16}$$

with the aid of ansatz (3). The requirement that *ρ* = 1, gives rise to a value of -2 for the kinetic energy per site in the limit *T* → 0, as indicated in Fig.4 for all values of *U/t* and *N*. In the limit *T* → ∞, one has for the plots in Fig.4(a) the limits −1(*N* = 2), −2/3(*N* = 3), −1/2(*N* = 4). For Fig.4(b), 0(*N* = 2), 2/3(*N* = 3), 1(*N* = 4). For Fig.4(c) 1(*N* = 2), 2(*N* = 3), 5/2(*N* = 4). Finally, for Fig.4(d), 8(*N* = 2), 34/3(*N* = 3), 13(*N* = 4).

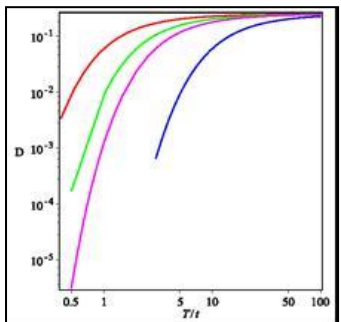

**Fig. 3a.** Number of on-site pairs *D* versus temperature for U/t = 4 (green), 8 (red), 12 (magenta), and 40 (blue) for N=2 with *ρ* = 1.

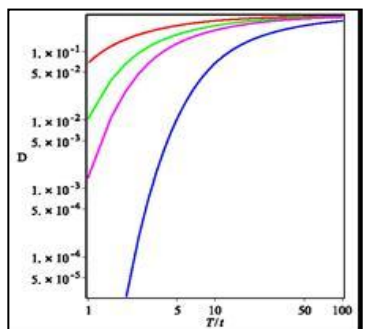

**Fig. 3b.** Same as Fig.3a but for N=3.

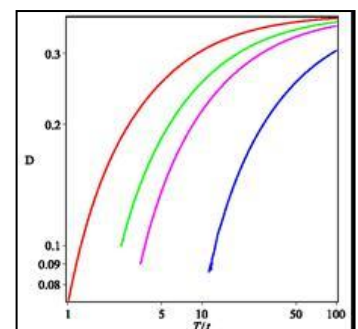

**Fig. 3c** Same as Fig.3a but for N=4.

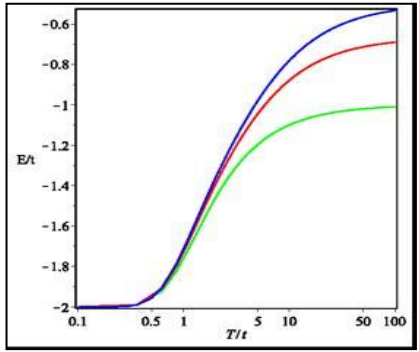

**Fig. 4a**. Energy per site E versus temperature T for N = 2 (green), 3 (red), 4 (blue) for $U/t = 4$ with $\rho = 1$.

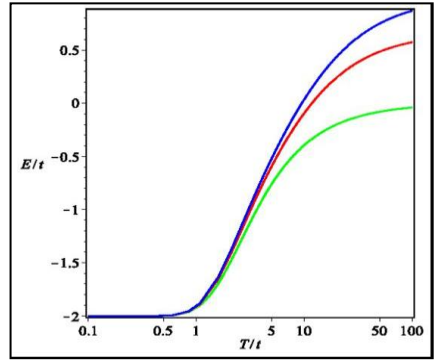

**Fig. 4b.** Same as Fig.4a but with $U/t = 8$.

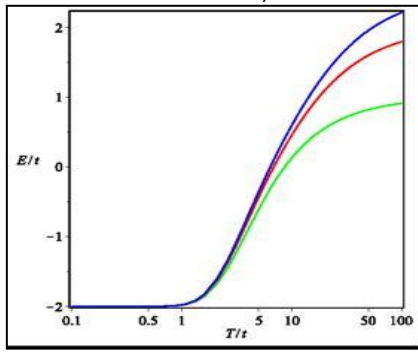

**Fig. 4c.** Same as Fig.4a but with $U/t = 12$.

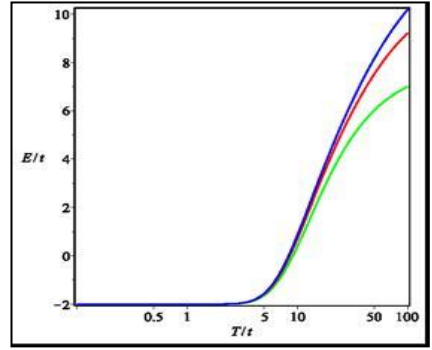

**Fig. 4d.** Same as Fig.4a but with $U/t = 40$.

The high and low temperature behaviors of the energy can be ascertained for the cases indicated in Fig.4, where the condition $\rho = 1$ has been imposed. However, when the total energy is expressed in terms of the renormalized $\tilde{\mu}$ and the scaled temperature $\tilde{T}$ one obtains, when the restriction $\rho = 1$ is not imposed, that

$$E \to \frac{1}{8}N(N+1)N_s \qquad (T \to \infty),$$
$$E \to \frac{1}{2}N^2 N_s \qquad (\tilde{\mu} \to \infty), \qquad (17)$$

depending only on N. The latter limit can be seen in Fig.10 of [2], where one has for $N = N_s = 6$, the limiting value of $E = 108$. The corresponding limit $T \to 0$ behavior will be discussed below, where the phase transition is made evident.

The specific heat is shown in Fig.5. Note that the location of the peaks are almost independent of N and increases with temperature as $U/t$ increases. Our results do not seem to agree with those in [3]. In Fig.6 one has that the entropy increases with increasing N for given $T/t$. The behavior of the entropy at low and high temperatures for the case with $\rho = 1$, differs considerably from cases where the requirement $\rho = 1$ is not imposed. In the former case one has

$$S/N_s = \ln N \qquad \text{at} \qquad T = 0$$
$$S/N_s = N \ln N - (N-1)\ln(N-1) \qquad \text{as} \qquad T \to \infty, \qquad (18)$$

as attested in Fig.6.

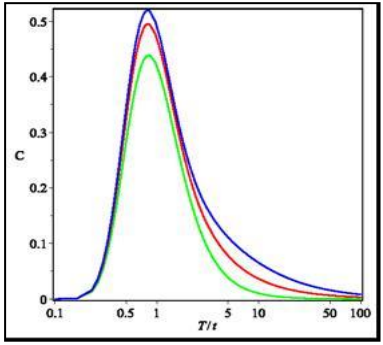

**Fig. 5a.** Specific heat versus temperature for N = 2 (green), 3 (red), 4 (blue) for *U/t* = 4 with *ρ* = 1.

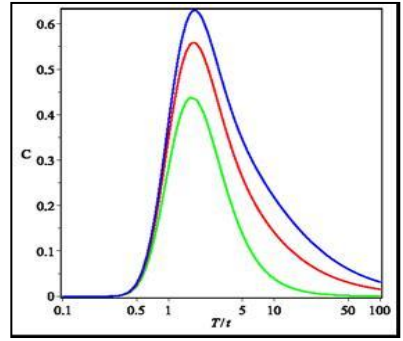

**Fig. 5b.** Same as Fig.5a but with *U/t* = 8.

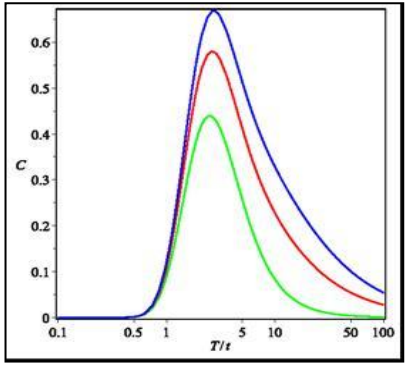

**Fig. 5c.** Same as Fig.5a but with *U/t* = 12.

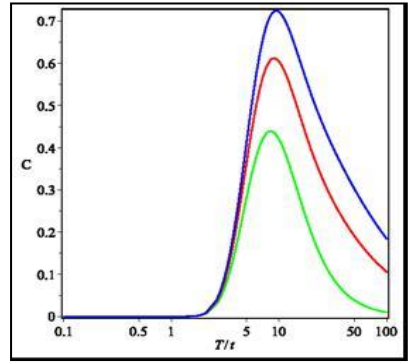

**Fig. 5d.** Same as Fig.5a but with *U/t* = 40.

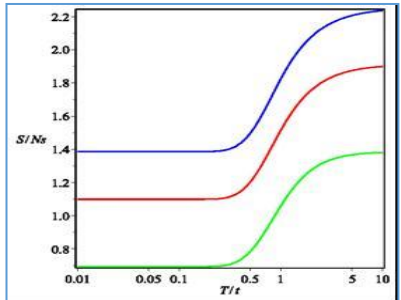

**Fig. 6a.** Entropy per site versus temperature for N = 2 (green), 3 (red), 4 (blue) for *U/t* = 4 with *ρ* = 1.

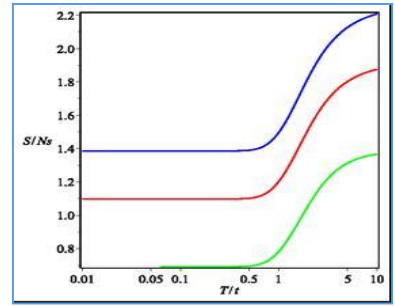

**Fig. 6b.** Same as Fig.6a but with *U/t* = 8.

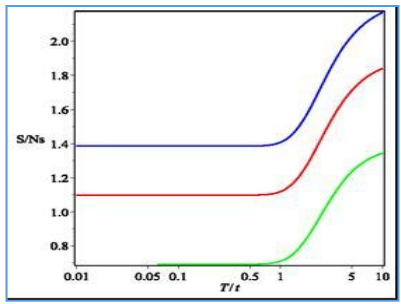

**Fig. 6c.** Same as Fig.6a but with *U/t* = 12.

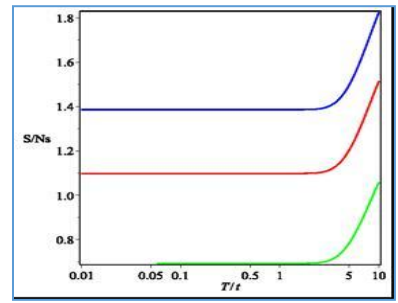

**Fig. 6d.** Same as Fig.6a but with *U/t* = 40.

In the general case where the condition *ρ* = 1 is not imposed, we obtain

$$S/N_s = \ln\left(\frac{4!}{(\frac{3}{4} + \frac{1}{8}\frac{\mu}{t})!(\frac{13}{4} - \frac{1}{8}\frac{\mu}{t})!}\right) \qquad \text{at} \qquad T = 0 \tag{19}$$

$$S/N_s = N \ln 2 \qquad \text{as} \qquad T \to \infty,$$

where *N* microstates in a given site have $2^N$ dimensions and so we can form *N* pairs of two-qubit entangled states each with a maximum entropy ln2 giving rise to a total entropy per site of *N*ln2. (see Fig.13 in [2], plot of total entropy with $N = N_s = 6$, which corresponds to our Fig.7(b) for the entropy per site $S/N_s$.). The values for *μ/t* indicated in the caption of Fig.7(a) corresponds to the values of the renormalized chemical potential, $\tilde{\mu} = 0, 1, 2, 3, 4$, where the constraint $\rho = 1$ is not imposed on the results. The plot in Fig.7(c) is the zero-temperature, scaled energy per site versus *N* for $\tilde{\mu} = 6$, where the constraint $\rho = 1$ is not imposed on the results,

$$E/(UN_s) = \frac{\tilde{\mu}^2}{2} \qquad \tilde{\mu} \leq N \quad \text{(constant)}$$
$$= \frac{N^2}{2} \qquad \tilde{\mu} > N \quad \text{(steps)}. \tag{20}$$

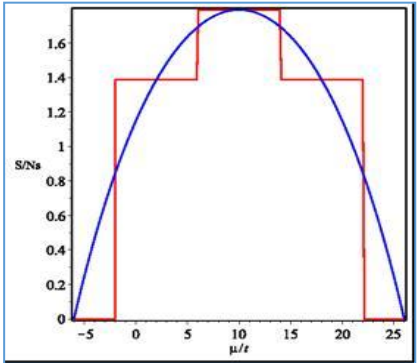

**Fig. 7a.** Zero temperature behavior of the entropy per site $S/N_s$ (red) and $\ln(N!/[\tilde{\mu}!(N-\tilde{\mu})])!$ (blue) versus *μ/t*, where $\tilde{\mu} = 2t/U + 1/2 + \mu/U$. For $N = 4$ lattice sites all with the same chemical potential and $U/t = 8$, $S/N_s = 0$ for $\mu/t = -6, 26$, $S/N_s = \ln(4) = 1.38...$ for $\mu/t = 2, 18$, $S/N_s = \ln(6) = 1.79...$ for $\mu/t = 10$.

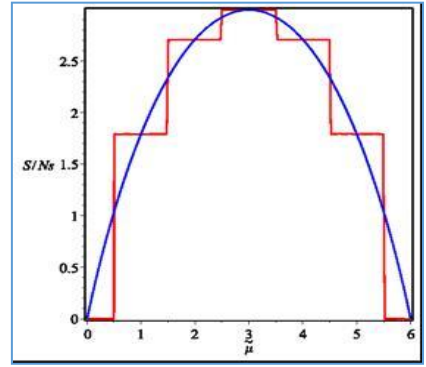

**Fig. 7b.** Zero temperature behavior of the entropy per site $S/N_s$ (red) and $\ln(N!/[\mu!(N-\mu)!])$ (blue) versus renormalized $\tilde{\mu} = 2t/U + 1/2 + \mu/U$ for $N = 6$ lattice sites all with the same chemical potential $\tilde{\mu}$. $S/N_s = 0$ for $0 < \tilde{\mu} < ½$ and $11/2 < \tilde{\mu}$, $S/N_s = \ln(6) = 1.79...$ for $1/2 < \tilde{\mu} < 3/2$ and $9/2 < \tilde{\mu} < 11/2$, $S/N_s = \ln(15) = 2.70...$ for $3/2 < \tilde{\mu} < 5/2$ and $7/2 < \tilde{\mu} < 9/2$, $S/N_s = \ln(20) = 2.99...$ for $5/2 < \tilde{\mu} < 7/2$.

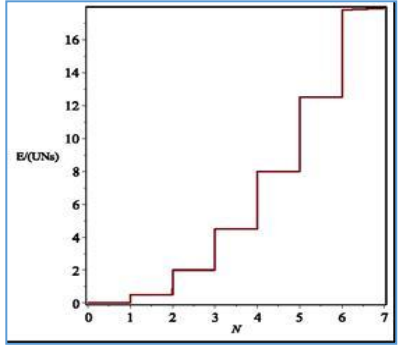

**Fig. 7c.** Zero temperature behavior of the energy per site $E/(UN_s)$ for $\tilde{\mu} = 6$ as a function of *N*. For $N \geq 6$ the energy is constant with value 18. The steps are for $N < 6$ corresponding to steps with height $N^2/2$.

Note that there is energy saturation for given renormalized chemical potential.

## 7. Conclusions

We consider further an exactly soluble, modified Fermi-Hubbard model in order to calculate thermodynamic observables and compare them to recently obtained numerical results that place the density constraint $\rho = 1$. It is not clear why the latter constraint; it may simplify the numerical computations. The modified model exhibits a continuous phase transition (second order) reminiscent of the integer quantum Hall resistance and a ground state, first-order phase transition. In the absence of said constraint, the high temperature behavior of the entropy per site shows maximal entanglement $N$ln 2, where $N$ is the number of microstates in a given lattice site.